\documentclass[preprint, 3p, twocolumn]{elsarticle}

\usepackage{lineno,hyperref}

\journal{Phys. Lett. B}

\DeclareGraphicsExtensions{.pdf}
\usepackage{amssymb,amsmath}
\graphicspath{{./image/}}

\begin{document}

\begin{frontmatter}

\title{Detecting the Upturn of the Solar $^8$B Neutrino Spectrum with LENA}

\author[cluster]{R. M\"{o}llenberg\corref{correspondingauthor}}
\cortext[correspondingauthor]{Corresponding author}
\ead{randolph.moellenberg@ph.tum.de}

\author[tum]{F. Feilitzsch}
\author[tum]{D. Hellgartner}
\author[tum]{L. Oberauer}
\author[tum]{M. Tippmann}
\author[mainz]{J. Winter}
\author[mainz]{M. Wurm}
\author[tum]{V. Zimmer}

\address[cluster]{Excellence Cluster Universe, Technische Universit\"{a}t M\"{u}nchen, 85748 Garching , Germany}
\address[tum]{Physik Department, Technische Universit\"{a}t M\"{u}nchen, 85748 Garching , Germany}
\address[mainz]{Institut f\"{u}r Physik, Excellence Cluster PRISMA, Johannes Gutenberg Universit\"{a}t Mainz, 55128 Mainz, Germany}

\begin{abstract}
LENA (\textbf{L}ow \textbf{E}nergy \textbf{N}eutrino \textbf{A}stronomy) has been proposed as a next generation
50\,kt liquid scintillator detector. The large target mass allows a high precision measurement of the solar $^8$B neutrino
spectrum, with an unprecedented energy threshold of 2\,MeV. Hence, it can probe the MSW-LMA prediction for the electron neutrino survival probability
in the transition region between vacuum and matter-dominated neutrino oscillations. 
Based on Monte Carlo simulations of the solar neutrino and the corresponding
background spectra, it was found that the predicted upturn of the solar $^8$B neutrino spectrum can be detected with $5\,\sigma$ significance after 5\,y.

\end{abstract}


\end{frontmatter}


\section{Introduction}


The MSW-LMA prediction for the survival probability of solar electron neutrinos $\mathrm{P_{ee}(E_\nu})$ has been confirmed
in the energy region of vacuum ($\mathrm{E_\nu \lesssim 1\,MeV}$) \cite{borexino_be7}
and matter dominated oscillations ($\mathrm{E_\nu \gtrsim 5\,MeV}$) \cite{borexino_b8,neutrinosk,neutrinosno}.
Nevertheless, the predicted upturn of the $^8$B neutrino spectrum in the transition region between vacuum and matter
dominated oscillations could not be detected, as Water-$ \mathrm{\check C}$erenkov detectors (WCDs) have a too high energy threshold and current liquid scintillator
detectors (LSDs) are too small.
A test of the MSW-LMA prediction in the transition region is important as new physiscs, like non-standard neutrino interactions \cite{NeutrinoNSI}
or light sterile neutrinos ($\mathrm{m_{\nu_1}<m_{\nu_0}<m_{\nu_2}}$, where the sterile neutrino $\mathrm{\nu_s}$ is mainly
present in the mass eigenstate $\nu_0$) \cite{light_sterile_neutrinos}, could influence $\mathrm{P_{ee}}$ in this region.

Compared to current solar neutrino detectors, the advantage of the proposed LENA detector \cite{lenawhitepaper} is the combination
of a $\sim$200\,keV energy threshold as well as a huge target mass. Thus, the external gamma background, which currently prevents a measurement of the solar $^8$B spectrum below 3\,MeV in Borexino \cite{borexino_b8}, can be suppressed by self-shielding.
This enables the measurement of the $^8$B spectrum with an unprecedented threshold of 2\,MeV.
Hence, LENA can probe the MSW-LMA prediction over a large part of the transition region.

The present work discusses the sensitivity of LENA to detect the predicted upturn of the $^8$B spectrum at low energies.
In Sec.\ \ref{sec:det} the planned detector setup is briefly presented. The simulation of the expected solar neutrino and background spectra is discussed
in Sec.\ \ref{sec:sim_nu} and Sec.\ \ref{sec:sim_bg}. The analysis of the simulated data is presented in Sec.\ \ref{sec:ana}. Finally,
the detection potential for the upturn of the solar $^8$B spectrum is discussed in Sec.\ \ref{sec:res}.

\section{The LENA Detector}
\label{sec:det}

The neutrino target consists of $\sim$50\,kt 
of liquid scintillator based on linear-akyl-benzene (LAB), that is enclosed in a cylinder with 14\,m radius
and 96\,m height \cite{lenaspec}. The emitted light is detected by photomultiplier tubes (PMTs)
that are mounted with non imaging light concentrators (LCs) inside individual pressure encapsulations
that are filled with a non-scintillating buffer liquid. The apertures of these optical modules are located at the boundary of the
target volume at a radius of 14\,m. The corresponding effective optical coverage is $\sim30\,\%$. The radius of the cylindrical concrete tank is 16\,m, so that the target volume is shielded by 2\,m of liquid scintillator.
A muon veto formed by gas detectors is placed above the detector tank and provides auxiliary information for the reconstruction of cosmic muon tracks.
In order to identify and reconstruct inclined muon tracks, an instrumented water volume surrounding the tank serves as an active 
Water-$ \mathrm{\check C}$erenkov muon veto and shields the target volume from fast neutrons.


The preferred location for the detector is the Pyh\"{a}salmi mine in Finland. The detector cavern is shielded by 1400\,m of rock coverage, corresponding to
$\sim$4000\,m water equivalent (w.e.). Hence, the cosmic muon flux will be reduced to $\sim\mathrm{0.2\,m^{-2}h^{-1}}$ \cite{fastneutronsim}, which
is about five times less than in Borexino.

\section{Simulation of the Solar Neutrino Spectra}
\label{sec:sim_nu}

There are two possible detection channels for solar $^8$B neutrinos in LENA. The elastic neutrino electron scattering (ES) channel
and charged current reactions of $\mathrm{\nu_e}$'s on $^{13}$C ($^{13}$C channel).

\subsection{Elastic Neutrino Electron Scattering}
In the ES channel, a neutrino scatters elastically off an electron, which is subsequently detected. As the energy of the recoil electron depends
on the scattering angle, the measured recoil spectrum is a convolution of the solar neutrino spectrum and the electron recoil spectrum at a given neutrino
energy.

The simulation of the electron recoil spectra was split into two parts. First of all, the differential events rates for neutrinos from different fusion reactions
were calculated acording to the BS05(AGS, OP) standard solar model \cite{standard_solar_model}. Using these differential events rates,
$10^6$ electron events were simulated with a GEANT4 \cite{geant4simtoolkit} based Monte Carlo (MC) simulation of the LENA detector \cite{moellenbergphd}.
The events were homogeneously distributed over the target volume, so that possible position dependent effects are considered. Afterwards, the visible energy
was reconstructed from the event position and the number of detected photons \cite{moellenbergphd}.

\begin{figure}[htbp]
\begin{center}
		
		\includegraphics[width=0.45\textwidth]{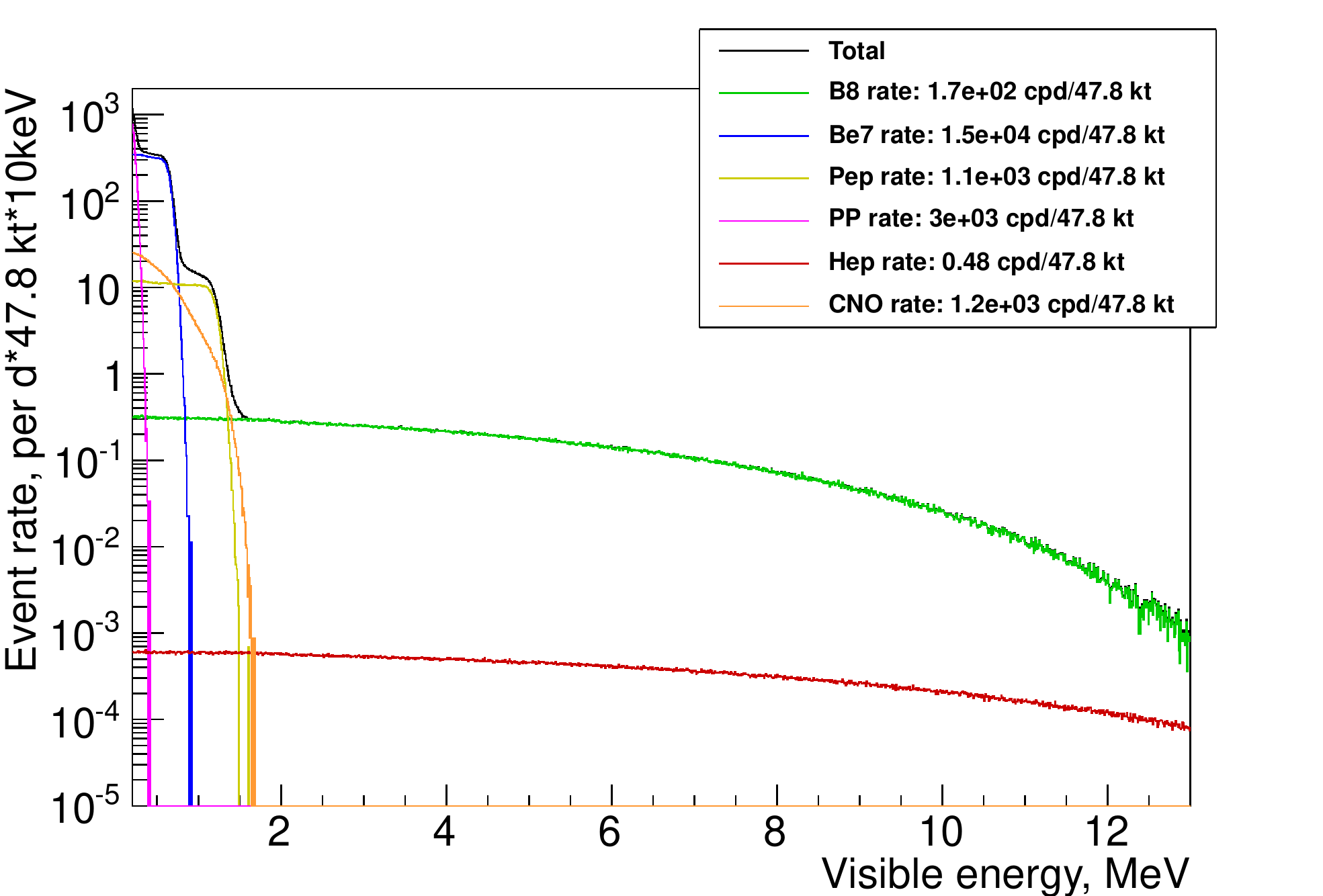}
\end{center}	
\caption[Solar neutrino spectra]{The simulated visible energy spectra for solar neutrinos with $\mathrm{P_{ee}(E_\nu)}$ according to the MSW-LMA prediction
for the ES channel.
The spectral shapes of the neutrino fluxes were taken from \cite{pp_spectrum,cno_spectrum,b8_spectrum,hep_spectrum}.}
\label{fig:solarspectra}
\end{figure}

Figure \ref{fig:solarspectra} shows the resulting electron recoil spectra.
At low energies, the $^7$Be and the pep neutrinos have the largest event rate, preventing a detection
of $^8$B neutrinos below $\sim1.4\,$MeV. Above $\sim1.4$\,MeV, the $^8$B spectrum is dominant and surpasses the hep spectrum by more than two orders
of magnitude.

\subsection{Charged Current Reaction on $^{13}$C}
The charged current reaction of electron neutrinos on $^{13}$C ($\mathrm{\nu_e+ {^{13}C} \rightarrow {^{13}N}+e^-}$) has a threshold
of 2.2\,MeV. Thus, it is another possible detection channel for solar $^8$B neutrinos. Due to the kinematics of the reaction, the recoiling $^{13}$N nucleus 
has only a few keV kinetic energy. Hence, the neutrino energy can be reconstructed on an
event-by-event basis by measuring the electron energy. The subsequent $\beta^+$ decay of the
$^{13}$N nucleus ($\tau=862\,$s) causes a delayed coincidence signal, which can be used to distinguish signal from background events. Hence,
$^{13}$C events will be selected by the spatial and timing coincidence between prompt and delayed signals.
Although the cross section is about 10 times
larger than the cross section of the ES channel \cite{nue_c13_xsec}, the event rate is almost two orders of magnitude
lower due to the low isotopic abundance of $^{13}$C (1.07\,\%).
Nevertheless, the measurement of the unconvoluted shape of the solar $^8$B neutrino spectrum with the $^{13}$C channel allows an energy dependent measurement of $\mathrm{P_{ee}}$, as the unoscillated solar $^8$B flux is known from the NC measurements of the SNO experiment \cite{sno_combined_analysis}.



\section{Simulation of the Background Spectra}
\label{sec:sim_bg}

There are three different types of background present in the ES channel: external gamma rays that are emitted by the tank, the PMTs and the LCs,
cosmogenic radioisotopes produced in-situ by traversing muons and intrinsic radioactive background.
A background for the $^{13}$C channel is caused by the accidental coincidences of these backgrounds and of ES interactions of solar neutrinos.

\begin{table}
\begin{center}
\begin{tabular}{|c|c|c|c|}
\hline
 &  $^{40}$K & $^{238}$U chain & $^{232}$Th chain \\
\hline
Tank & 13\,MBq & 1.1\,GBq & 178\,MBg\\
PMTs & 14\,kBq & 229\,kBq & 24\,kBg\\
LCs & 0.86\,kBq & 13\,kBq & 41\,kBg\\
\hline
\end{tabular}
\end{center}
\caption[Gamma rates]{The gamma rates above 250\,keV of the different detector components.}
\label{tab:gamma_rates}
\end{table}

Based on the assumed radiopurity of the tank \cite{lenaspec}, PMTs and LCs \cite{maneira_pmt} (see Table
\ref{tab:gamma_rates}), the external gamma ray background was simulated with the GEANT4-based 
LENA Monte Carlo simulation \cite{moellenbergphd}. It was found that no external gamma background is present above 3.5\,MeV.
For lower energies, the rate can be reduced to a negligible level by applying a fiducial volume cut. The correspindig fiducial volume is 48\,kt above
and 19\,kt below 3.5\,MeV.

Cosmogenic radioisotopes are produced inside the target volume by spallation reactions of cosmic muons on carbon nuclei. The majority 
of the produced radioisotopes have a lifetime of less than $\sim 1\,$s \cite{borexino_b8, KamLANDCosmogenics}.
Hence, the decays can easily be identified by the time coincidence to the parent muon, without introducing a large dead time.
The remaining cosmogenic isotopes with a longer lifetime are $^{11}$C ($\beta^+$), $^{10}$C ($\beta^+$) and $^{11}$Be ($\beta^-$)
(see Table \ref{tab:cosmogenic_isotopes}).
The spectral shapes of these isotopes were obtained from the GEANT4-based LENA Monte Carlo simulation. Afterwards, the measured rates of the Borexino
experiment\cite{borexino_b8, borexino_be7} have been scaled to the Pyh\"{a}salmi location, using the muon flux of the two sites\footnote{Note that no scaling
for the slightly different mean muon energy was applied.}.
Below 2\,MeV, the $^{11}$C background is about two orders of magnitude larger than the solar $^8$B neutrino signal. Hence, the end of the $^{11}$C spectrum
defines the energy threshold for the detection of solar $^8$B neutrinos.
As $^{10}$C and $^{11}$Be have a much shorter lifetime than $^{11}$C, it is possible to reduce the background from these isotopes by
vetoing a cylinder with 2\,m radius around each traversing muon for $\mathrm{\Delta t=4\cdot \tau (^{10}C)=111.2\,s}$. 
As the muon rate in the fiducial volume is  $\mathrm{\sim135\,h^{-1}}$,
the introduced dead time amounts to about 10\,\% of the total exposure, which is still acceptable.

\begin{table}
\begin{center}
\begin{tabular}{|c|c|c|c|c|}
\hline
Isotope& Q-Value & Life time &Rate [cpd/\,kt]\\
\hline
$^{11}$C& 2.0\,MeV  & 29.4\,min & 54\\
$^{10}$C& 3.7\,MeV & 27.8\,s & 1.0\\
$^{11}$Be& 11.5\,MeV & 19.9\,s & $6.4\cdot10^{-2}$\\ 
\hline
\end{tabular}
\end{center}
\caption[Cosmogenic radioisotopes]
{List of the cosmogenic radioisotopes with life times above 2\,s.}
\label{tab:cosmogenic_isotopes}
\end{table}

\begin{figure}[!htbp]\centering\includegraphics[width=0.48\textwidth]{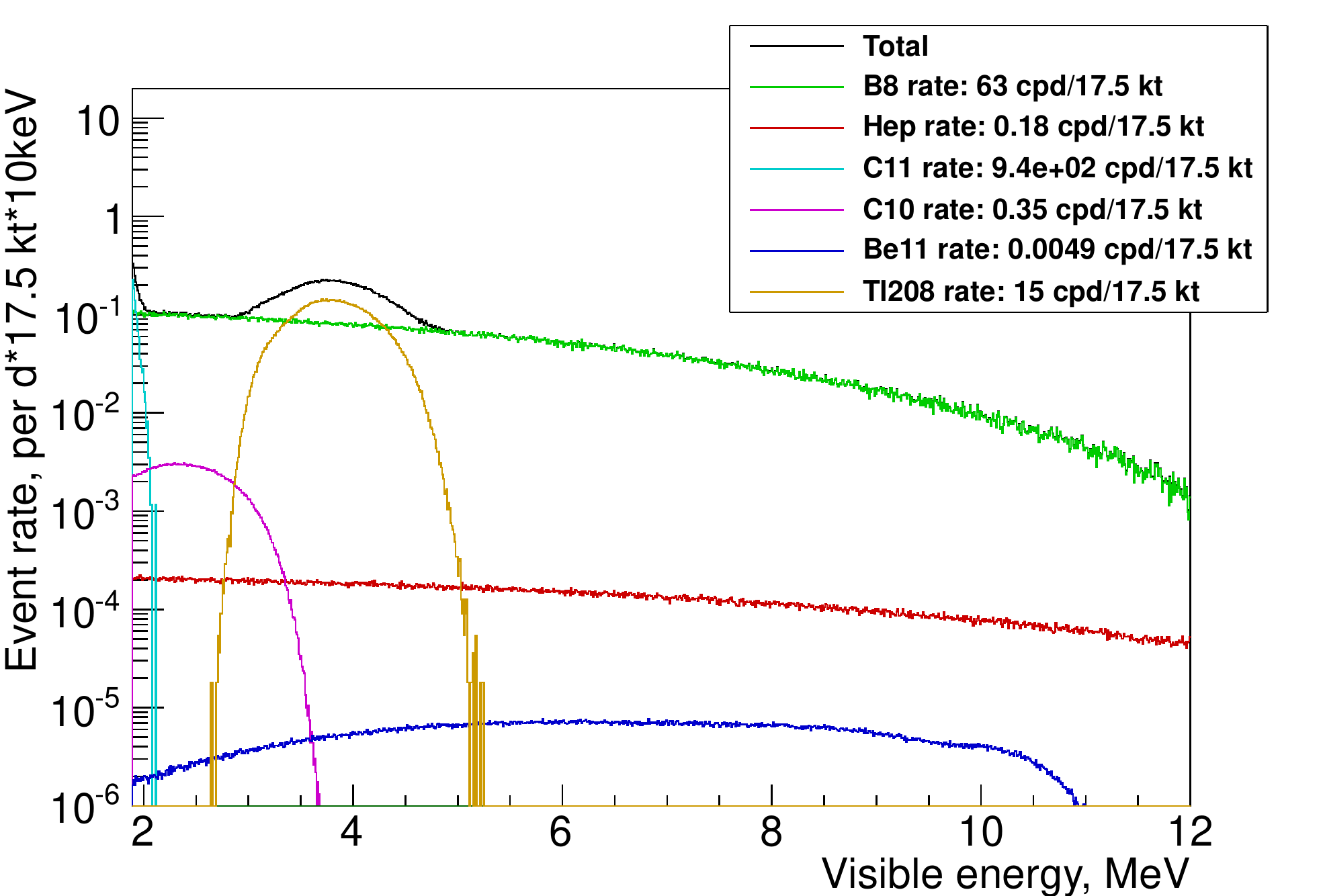} 
\caption[Background spectra (ES channel)]{The simulated visible energy spectra of the cosmogenic and intrinsic radioactive background.
A spacial and time cut around each muon was applied to reduce the cosmogenic background and
the external gamma background was suppressed to a negligible level by a fiducial volume cut.}
\label{fig:background_es}
\end{figure}

Besides cosmogenic isotopes, there is also a background from intrinsic radioimpurities in the scintillator. As the amount of radioimpurities in the LENA
detector is not known at the moment, it was assumed in the following that the radiopurity levels of the Borexino experiment\footnote{Note that the radiopurity levels of the first data taking phase of Borexino are used and that
the current radioactive background rates in Borexino have been substantially improved by several purification campaigns.} ($\mathrm{c(^{232}Th)=(6.5\pm1.5)\cdot10^{-18} g/g}$) \cite{borexino_b8} are reached.
The only intrinsic beta-emitting\footnote{Alpha emitters above 2\,MeV do not pose a background as their light emission is quenched in a liquid scintillator.} radioisotopes in the Borexino detector with a Q-Value above 2\,MeV are $^{214}$Bi ($^{238}$U chain, Q=3.3\,MeV)
and $^{208}$Tl ($^{232}$Th chain, Q=5.0\,MeV) \cite{borexino_b8}. $^{214}$Bi can be tagged by the subsequent decay of $^{214}$Po and is thus neglected in the following \cite{borexino_b8}. $^{208}$Tl is produced by the alpha decay of $^{212}$Bi which
also decays into $^{212}$Po ($\tau=0.4\mu s$) with 64\,\% branching ratio.
Hence, the amount of $^{208}$Tl can be determined from the observed number of $^{212}$Bi-$^{212}$Po
concidences. Figure \ref{fig:background_es} shows the resulting cosmogenic and intrinsic radioactive background spectra.

\begin{figure}[!htbp]\centering\includegraphics[width=0.48\textwidth]{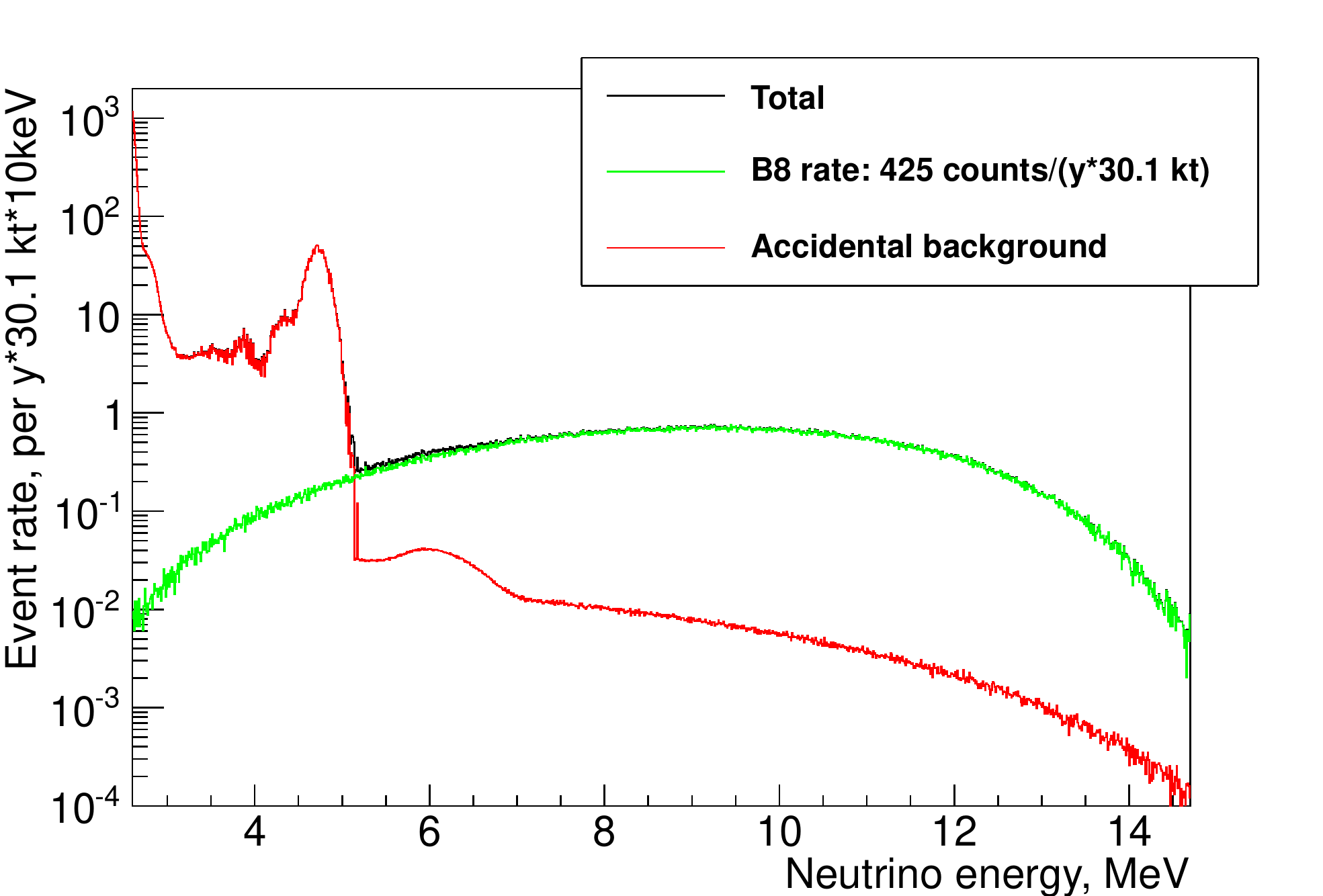} 
\caption[The accidental background spectrum for the $^{13}$C channel]{The accidental background spectrum for the $^{13}$C channel, using a fiducial
volume with 11\,m radius. Furthermore, the expected $^8$B spectrum is shown for comparison.}
\label{fig:acc_background}
\end{figure}

The accidental background for the $^{13}$C channel has been calculated (see Figure \ref{fig:acc_background}) based on these spectra. 
Due to the relatively long life time of the $^{13}$N nucleus, a large amount of background is present below 5\,MeV reconstructed neutrino energy.
Above 5\,MeV, the accidental background is at least one order of magnitude below the solar $^8$B signal. Thus, the energy
threshold for the $^{13}$C channel is set to 5\,MeV neutrino energy.

\section{Analysis Procedure}
\label{sec:ana}

\begin{figure}[!htbp]\centering\includegraphics[width=0.48\textwidth]{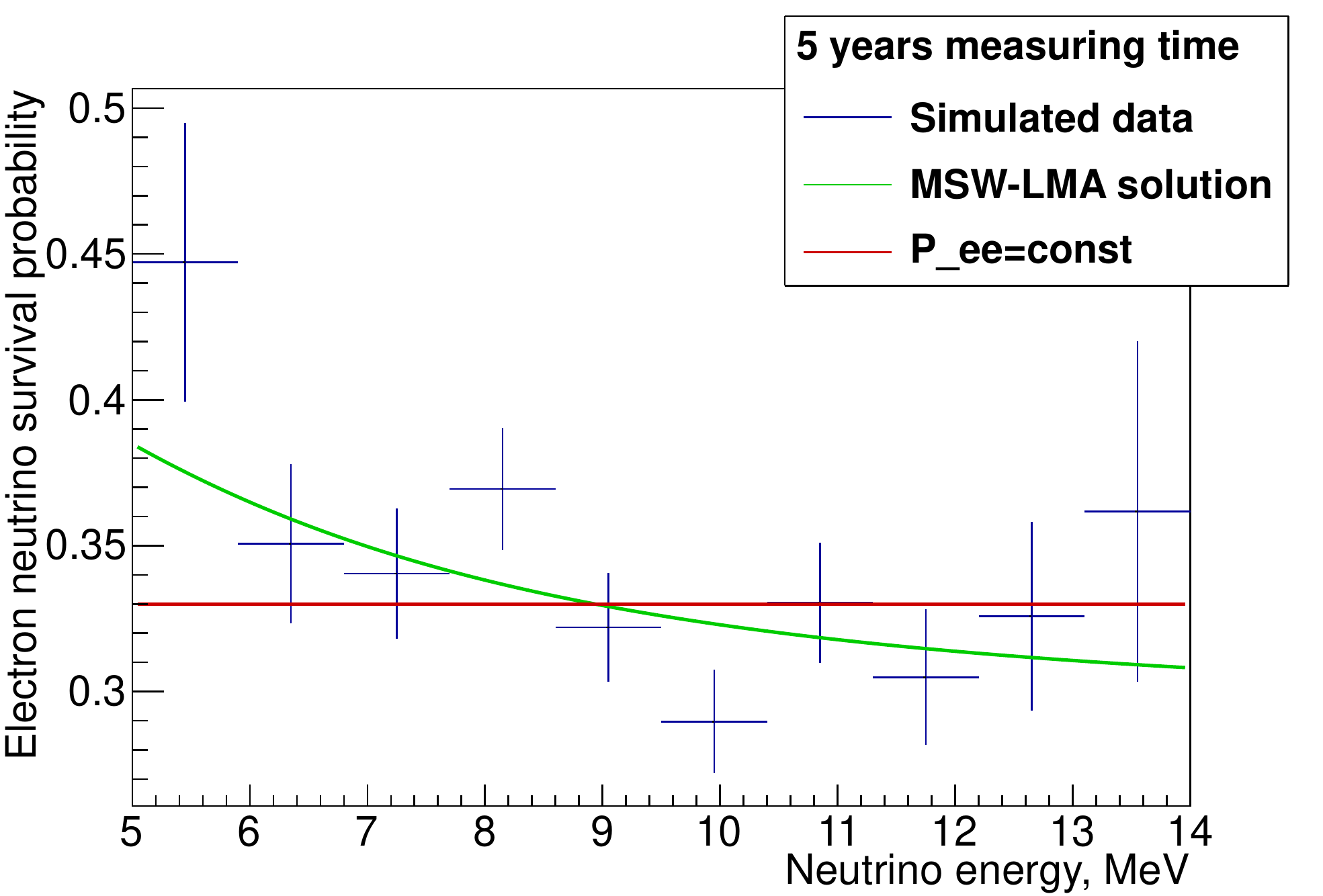} 
\caption[The measured $\mathrm{P_{ee}}$ ($^{13}$C channel)]{
The measured electron neutrino survival probability after 5\,years, using  the $^{13}$C channel. Furthermore, a fit with the MSW-LMA prediction (depicted in green)
and with $\mathrm{P_{ee}=const}$ (depicted in red) is shown.}
\label{fig:pee_example_c13}
\end{figure}

In order to determine the sensitivity for the detection of the upturn of the solar $^8$B neutrino spectrum, the potential to distinguish the MSW-LMA prediction from a simple test model with $\mathrm{P_{ee}(E_\nu)=const}$ was investigated.
Using the previously simulated neutrino and background spectra,
$10^5$ five year long measurements of the $^{8}$B neutrino spectrum were simulated with the ROOT package \cite{root}.
Afterwards, the MC data sets were analyzed independently for the $^{13}$C channel and the ES channel.
The accidental background was subtracted from the $^{13}$C channel to retrieve the oscillated solar $^8$B spectrum. Afterwards, this spectrum was divided by the unoscillated solar $^8$B spectrum to determine $\mathrm{P_{ee}(E_\nu)}$. Note that the accidental background spectrum can be precisely determined from the
measured total spectrum of the ES channel.
Finally, $\mathrm{P_{ee}(E_\nu)}$ was fitted with the MSW-LMA prediction and with
the $\mathrm{P_{ee}(E_\nu)=const}$ model, using a $\chi^2$ minimization.
The normalization was treated as a free nuisance parameter in both cases.
Hence, this analysis is only sensitive to the shape of $\mathrm{P_{ee}(E_\nu)}$ and is thus unaffected by the uncertainty of the solar $^8$B flux.
Figure \ref{fig:pee_example_c13} shows the result for one example measurement. While the MSW-LMA prediction is preferred one, the statistical 
significance is not enough to exclude the $\mathrm{P_{ee}(E_\nu)=const}$ model. Overall, the average exclusion significance in the
$^{13}$C channel is below $1\,\sigma$.

\begin{figure}[!htbp]\centering\includegraphics[width=0.48\textwidth]{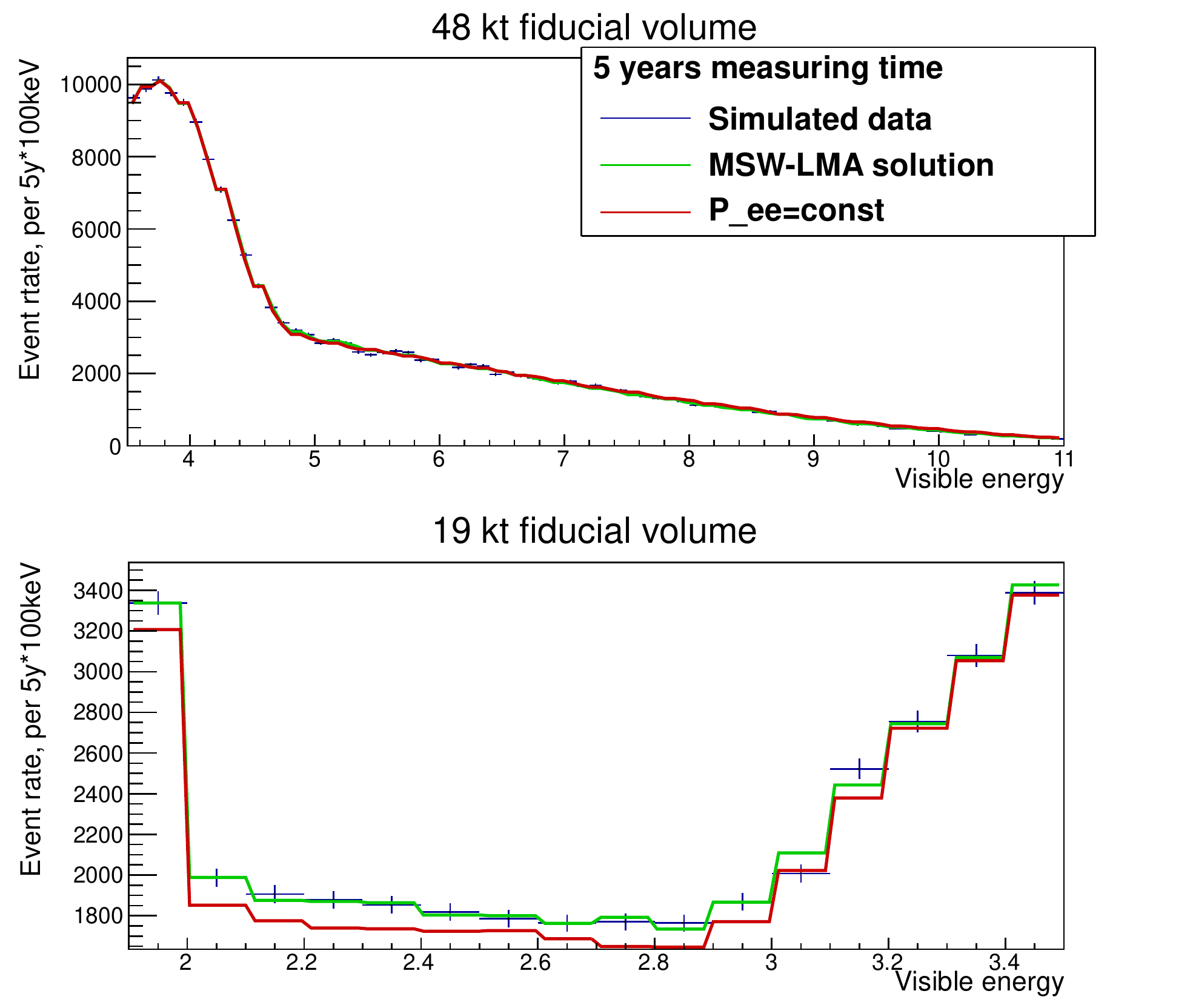} 
\caption[The simulated total spectrum for the ES channel]{
The total spectrum for the ES channel after 5\,y measuring time. Furthermore, a fit according to the MSW-LMA prediction (depicted in green)
and according to $\mathrm{P_{ee}=const}$ (depicted in red) is shown.}
\label{fig:fit_example_es}
\end{figure}

For the ES channel, it is not possible to directly calculate $\mathrm{P_{ee}}$ from the measured $^8$B spectrum, as the neutrino energy can not be reconstructed
on an event-by-event basis. Thus, the simulated spectrum was fitted with the expected spectrum according to the MSW-LMA prediction and
the $\mathrm{P_{ee}(E_\nu)=const}$ hypothesis, using a $\chi^2$ minimization with free parameters
for the solar neutrino and the corresponding background rates. In order to maximize the statistics, the full 48\,kt fiducial volume was used 
above 3.5\,MeV, as no external gamma background is present at these energies, while the 19\,kt fiducial volume was used below 3.5\,MeV. As the contributions of the cosmogenic and intrinsic radioactive background
can be measured independently, pull-terms were added to the $\chi^2$ function to maximize the sensitivity \cite{kathrinphd}:

\begin{equation}
\mathrm{\chi^2_{tot}=\chi^2+\chi^2_{pull}}, \quad  \mathrm{\chi^2_{pull}=\sum_{j=1}^{k} \frac{\left( \lambda_j-\mu_j \right)^2}{\sigma_{\lambda_j}^2} \ , }
\end{equation}

where k is the number of parameters with prior information, $\lambda_j$ is the fit value of the parameter j and $\mu_j$ is the expected value of the parameter j, with
$\sigma_{\lambda_j}$ uncertainty.
The uncertainty for the cosmogenic backgrounds were taken from the KamLAND ($^{10}$C and $^{11}$Be) \cite{KamLANDCosmogenics} and the 
Borexino experiment ($^{11}$C) \cite{borexino_be7}, while the uncertainty for the $^{208}$Tl rate was estimated from the expected number of
$^{212}$Bi-$^{212}$Po coincidences. Figure \ref{fig:fit_example_es} shows the results of the fit for one example measurement. Above $\sim 3\,$MeV visible energy, the data is consistent with both the MSW-LMA prediction and with the $\mathrm{P_{ee}=const}$ hypothesis. But below $\sim 3\,$MeV, the MSW-LMA prediction is clearly favored, which shows the importance of measuring the $^8$B spectrum below 3\,MeV, which is not possible with current WC and LS detectors.

In the last step, the results from the $^{13}$C and the ES channel were combined by adding the $\chi^2$ values of the corresponding fits. Using these
values and the number of degrees of freedom, the probability that the MC data sample is consistent with the MSW-LMA prediction or the $\mathrm{P_{ee}(E_\nu)=const}$
model was calculated. In order to suppress statistical fluctuations, this process was repeated for each data sample. Finally,
the probability that the $\mathrm{P_{ee}(E_\nu)=const}$ model can be excluded with $5\,\sigma$ significance, which is equivalent to a detection of 
the upturn of the $^8$B spectrum, was calculated, assuming that the MSW-LMA prediction is correct.

\section{Results}
\label{sec:res}

Table \ref{tab:const_excl} shows the probability for a $5\,\sigma$ detection of the upturn of the solar $^8$B neutrino spectrum, as a function
of the measurement time.
After 2 years, the upturn can be detected at $5\,\sigma$ significance for over 40\,\% of all MC data sets, assuming
that the MSW-LMA prediction is correct. After 5\,y, the upturn was detected for each data set.
Hence, in case that the upturn is not detected after 5\,y, the MSW-LMA prediction
would be ruled out and new physics must be present that reduce $\mathrm{P_{ee}(E_\nu)}$ in the transition region.
Comparing the two detection channels, it was found that the sensitivity of the ES channel is much larger than the $\mathrm{^{13}C}$ channel, due 
to the larger statistics. Nevertheless, the $\mathrm{^{13}C}$ channel still provides an important cross check of the results.

\begin{table}
\begin{center}
\begin{tabular}{|c|c|}
\hline
measuring time  &prob. for a 5\,$\sigma$ det.\\
\hline
2 years & 43.4\,\% \\
3 years & 92.5\,\%\\
4 years & 99.8\,\%\\
5 years & $>99.9$\,\%\\
\hline
\end{tabular}
\end{center}
\caption[Detection potential for the $^8$B upturn]
{The probability to detect the upturn of the solar $^8$B spectrum, for measuring times ranging
from 2\,y to 5\,y.}
\label{tab:const_excl}
\end{table}

While the amount of cosmogenic background can be precisely estimated for the assumed rock coverage, it is much harder
to estimate the intrinsic radioactive background. Hence, the analysis was repeated for a 100 times larger instrinsic
radioactive background than in Borexino. Table \ref{tab:const_excl_100xBg} shows the detection potential for the upturn
of the solar $^8$B neutrino spectrum in this pessimistic scenario. While the detection potential is of course decreased, the effect
of the increased background is not very strong and the upturn can still be detected at $5\,\sigma$ significance after 5\,y.
The reason for this behaviour is that the important energy region below 3\,MeV is not affected by the larger $^{208}$Tl background.
Hence, a precision test of the MSW-LMA prediction is possible with LENA even if radiopurity conditions are substantially worse than in Borexino.

\begin{table}
\begin{center}
\begin{tabular}{|c|c|}
\hline
measuring time &prob. for a 5\,$\sigma$ det.\\
\hline
2 years & 34.6\,\% \\
3 years & 86.7\,\%\\
4 years & 99.4\,\%\\
5 years & $>99.9$\,\%\\
\hline
\end{tabular}
\end{center}
\caption[Detection potential for the $^8$B upturn for a two orders of magnitude larger instrinsic radioactive background]
{The probability to detect the upturn of the solar $^8$B spectrum, for measuring times between
2\,y and 5\,y and for a 100 times larger intrinsic radioactive background than in Borexino.}
\label{tab:const_excl_100xBg}
\end{table}

\section{Conclusions}
\label{sec:con}

Present-day experiments lack the capability for a precision measurement  of the electron neutrino survival probability $\mathrm{P_{ee}}$ in the transition region between vacuum and matter dominated
oscillations ($\mathrm{1\,MeV \lesssim E_\nu \lesssim 5\,MeV}$). 
LENA will offer an excellent opportunity to close this gap in the determination of $\mathrm{P_{ee}}$ by a high-statistics, low-energy-threshold measurement
of the solar $^8$B neutrino spectrum.
Due to its large target mass, the external gamma background, that currently prevents a measurement below 3\,MeV electron recoil energy in Borexino, can be efficiently
suppressed by a stringent fiducial volume cut. This allows a measurement of the solar $^8$B neutrino spectrum with an unprecedented energy threshold of 2\,MeV.

In the present work, the detection potential for the spectral upturn of the solar $^8$B spectrum that is predicted by the MSW-LMA solution was analyzed.
It was found that the spectral upturn can be detected at $5\,\sigma$ significance after 5\,y measuring time,
even if the intrinsic radiopurity level of the scintillator is two orders of magnitude worse than achieved in Borexino.
In case that the upturn of the solar $^8$B neutrino spectrum is not found, the measurement would rule out the MSW-LMA prediction and show that new physics decrease $\mathrm{P_{ee}}$ in the transition region between vacuum and matter dominated oscillations. 


\section*{Acknowlegdements}

This research was supported by the DFG cluster of excellence 'Origin and Structure of the Universe' (Munich) and 'PRISMA' (Mainz).

\bibliography{lit}
\end{document}